# Indoor Localization System of ROS mobile robot based on Visible Light Communication

Weipeng Guan[1,#], Shihuan Chen[3,#], Shangsheng Wen[2], Wenyuan Hou[2],

Zequn Tan[3], and Ruihong Cen[4]

1. Department of Information Engineering, The Chinese University of Hong Kong, Shatin, New Territories, Hong Kong, China
2. School of Materials Science and Engineering, South China University of Technology, Guangzhou 510640, China
3. School of Automation Science and Engineering, South China University of Technology, Guangzhou 510640, China
4. School of Mechanical and Automotive Engineering, South China University of Technology, Guangzhou, Guangdong 510640, China

#These authors contributed equally to this work, Guan Weipeng and Chen Shihuan are co-first authors of the article

Corresponding author: Shangsheng Wen (shshwen@scut.edu.cn), Weipeng Guan (gwpscut@163.com, weipengguan@cuhk.edu.hk)

**Abstract:** In this paper, an indoor robot localization system based on Robot Operating System (ROS) and visible light communication (VLC) is presented. On the basis of our previous work, we innovatively designed a VLC localization and navigation package based on Robot Operating System (ROS), which contains the LED-ID detection and recognition method, the video target tracking algorithm and the double-lamp positioning algorithm. This package exploited the principle of double-lamp positioning and the loose coupling characteristics of the ROS system, which is implemented by loosely coupled ROS nodes. Consequently, this paper combines ROS and VLC, aiming at promoting the application of VLC positioning in mature robotic systems. Moreover, it pushed forward the development of localization application based on VLC technology and lays a foundation for transplanting to other ROS robot platforms. Experimental results show that the proposed system can provide indoor localization within 1 cm and possesses a good real-time performance which takes only 0.4 seconds for one-time positioning. And if a high-performance laptop is used, the single positioning time can be reduced to 0.08 seconds. Therefore, this study confirms the practical application and the superior performance of VLC technology in ROS robot, showing the great potential of VLC localization. The video demo of the proposed robot positioning system based on VLC can be seen in *.

**Index Terms:** indoor positioning, Optoelectronic/photonic sensors, Robot Operating System (ROS), TurtleBot3, visible light communication, visible light positioning.

1. Introduction
1.1. VISIBLE LIGHT COMMUNICATION

Visible light wireless communication, also known as "light fidelity technology" (LiFi) is a wireless transmission technology that uses visible light spectrum for data transmission. It uses electrical signals to control the high-speed flashing LEDs to transmit information. LED always appears in light-belt decoration, advertising signs, streetlights, as well as indoor lighting [1]. For its unique strength like durability, controllability, high cost-effectiveness, long life-expectancy and environmental-friendly characteristic, LEDs are regarded as the main lighting devices [2]. What is more, the modulation frequency of LEDs is so high (>300 MHz) that it is undetectable by the human eyes, yet can be detected by photodiodes (PDs) and Complementary Metal Oxide Semiconductor (CMOS) cameras, which enable LEDs to be used for communication [1,3]. Hence, Visible Light Communication (VLC) technology comes into being. LED based VLC systems using PDs, cameras and linear arrays have been introduced in [3-6]. Relying on the existing LEDs, Visible Light Communication integrates lighting and communication, which can cover all the corners reached by the LED light. There are already various applications of VLC system such as the link between vehicles and traffic lights [7] and indoor information broadcasting [8, 9, 1]. The wide application and industrial upgrade of LEDs has also created conditions for VLC to compete with traditional radio frequency (RF) wireless communications. Moreover, the visible light band is still a blank spectrum and can be used without authorization. Therefore, VLC technology has rich spectrum resources and little external interference, which can expand the spectrum of the next generation broadband communication technology.

---

*Demonstration video is available at: https://kwanwaipang.github.io/Image/ROS.mp4

## 1.2. VISIBLE LIGHT POSITIONING

Visible light positioning is a positioning technology based on visible light communication. It is a good solution for the urgently needed indoor positioning due to the boom in the number of large buildings. Although in the past few years, the satellite positioning systems including Global Positioning System (GPS), the BeiDou Navigation Satellite System (BDS) and the Global Navigation Satellite System (GLONASS), has been rapidly developed. Nowadays, with the help of some professional equipment and base stations, they can even provide centimetre-level positioning services based on Real-time kinematic (RTK) technology, which can satisfy the need of outdoor positioning. However, as satellite signal cannot penetrate the walls [10], indoor positioning is still a problem to be solved. Many indoor positioning technologies are being proposed and used, such as Wi-Fi, Bluetooth, RFID, [11-13] and inertia navigation, but unfortunately, they are still far away from the ideal high-precision positioning in daily life due to theirs low accuracy, heave cost and non-portability etc. However, the visible indoor positioning can handle this problem well. visible light positioning (VLP) is based on visible light communication (VLC), which employs the LED lamps as the transmitter to transmit the ID information and uses the photoelectric device or the camera as positioning terminals to realize optical communication and indoor positioning. Compared with the traditional indoor positioning technology, the VLC based positioning technology has the advantages of high positioning accuracy, no electromagnetic interference, and environmental-friendly characteristic, thus having aroused the attention of many experts and scholars in the world.

In general, VLP indoor positioning has two kinds of physical components to choose: Photodetector (PD) [1] or image sensor [4]. The PD-based VLC positioning system is simple to implement, which does not require image processing technologies, so the cost is relatively low. But as studies have shown [14], the accuracy of PD-based VLC positioning system depends to a large extent on the direction of the beam, which is not ideal for indoor location. In addition, its positioning result is inaccurate as the measured signal strength varies with the intensity of the ambient light, resulting in unacceptable errors in many positioning results. Although the VLC positioning technology based on image sensor requires corresponding image processing technology and is limited by the angle of view of the image sensor, it is less affected by the system and the outside world, and the positioning accuracy is higher than that of the PD-based VLC positioning scheme. Furthermore, with the popularity of image sensors installed on intelligent terminals such as wearable devices and robots, VLC positioning technology based on image sensors has broader prospect and more commercial potential than that based on PD receiver. Therefore, the focus of this paper is on image sensor based VLC positioning technology.

**1.3. ROBOT INDOOR POSITIONING**

With the development of intelligent technology, robots are gaining more and more popularity in many areas, making robot indoor positing technology more important than ever. Over the years, the localization of mobile robots has been well studied, meeting the conditions of positioning even in industrial environments where high positioning accuracy are required. Dynamic estimation method where the position information depends on the outputs of the controller is one of the most popular estimation methods [15]. However, only relying on noumenal sensing is easy to have drift, having a bad impact on the autonomy-level of industrial task, which requires localization in the environment for a long time without interruption. Beyond that, WiFi-based positioning is also a common solution for indoor positioning. But owing to its inherent defect, WiFi-based indoor positioning has been limited to precision problems. Even if the data fusion of multiple access points is carried out through particle filtering, the positioning error is still above 1m [16]. Another framework for robot platforms based on image and laser sensors is presented in [17], in which the accuracy of the two positioning methods are analyzed, and Monte Carlo Localization (MCL) algorithm and CNN algorithm are used to improve the positioning accuracy and robustness.

All above, some common methods for indoor positioning with their own inherent shortcomings are demonstrated. Looking to the future, the indoor application of the robots is closely related to the precise indoor positioning technology, but the existing positioning technologies such as Dynamic estimation method, Wi-Fi, image-based method and so on still yet provide satisfying positioning accuracy. In this paper, we proposed to apply visible light positioning (VLP) to indoor robot positioning. Based on visible light communication and already widely installed LED lamps, visible light position can be a low-cost, high-precision and easy-to-deploy solution for robot indoor positioning.

**1.4. CONTRIBUTIONS AND OUTLINE**

To the best of our knowledge, there are not similar attempts to apply visible light communication (VLC) to the ROS before. Although there are a range of indoor VLP researches, as theoretical study shown in [20, 3, 21, 22, 23], they are not practical solutions but basic applications. Other researches of indoor positioning use microcontroller such as STM32 to achieve it. As is shown in [24, 25, 26, 4], the characteristic of attaching to a specific set of hardware hinder their software to migrate to other platforms. As a result, in this paper, we developed a VLC program package based on Robot Operating System (ROS). Aiming at promoting the application of VLC positioning in mature robotic systems, this package combines ROS and VLC, which pushed forward the development of localization application based on VLC technology and lays a foundation for transplanting to other ROS robot platforms. Further experiment tested its validity, which shows that the proposed system can provide indoor localization within 1 cm and possesses a good real-time performance which takes only 0.4 seconds for one-time positioning. And the single positioning time can be reduced to 0.08 seconds if a high-performance laptop is used. Therefore, this study confirms the practical application and the superior performance of VLC technology in ROS robot, showing the great potential of VLC localization for indoor robot positioning.

The principles of the visible light positioning (VLP) and Robot Operating System (ROS) architecture are shown in system principle and the experimental results are given in the section experiments.

## 2. System Principle
### 2.1. THE PRINCIPLE OF VISIBLE LIGHT POSITIONING

In this paper, VLC is exploited for indoor robot positioning which based on a double-lamp experiment platform and a TurtleBot. In our previous research, in order to solve the problems of traditional CMOS-based LED-ID modulation and demodulation methods, such as low communication efficiency, poor anti-interference ability, few lamp addresses, short distance, etc., we proposed a LED-ID optical stripe-code modulation and recognition algorithm based on VLC, which gives every LED unique features and matches them with the exact location information. By comparing the extracted and expected LED-ID, measuring would be straightforward. Furthermore, we have innovatively applied the Camshift algorithm to Visible Light Positioning (VLP) and proposed a Kalman Filter Tracking Algorithm based on improved Camshift Algorithm, using video target tracking algorithm to achieve dynamic positioning. Based on the accurate identification of LED-ID and the dynamic tracking detection of Camshift algorithm, a high-precision VLC imaging positioning algorithm based on two LED lamps is proposed to obtain the accurate three-dimensional coordinates of the object. This experiment also considers the accuracy, real-time and robustness of the VLC positioning system.

### 2.1.1. The LED-ID Detection and Recognition Method Based on Visible Light Communication

To solve the low positioning accuracy and poor real-time ability of traditional VLC positioning technologies relying on the angle of arrival (AOA) of visible light or the rich site summary (RSS) of light to calculate the distance between each target [27], as shown in Fig. 1, we exploited the Visible Light Imaging Positioning technology in this experiment to obtain the coordinates of the terminal. The CMOS image sensors are chosen as the visual sensor of the robot system, which uses Rolling Shutter mechanism where the data are exposed and read out row by row and then merged together to form an image [5, 28, 30, 31].

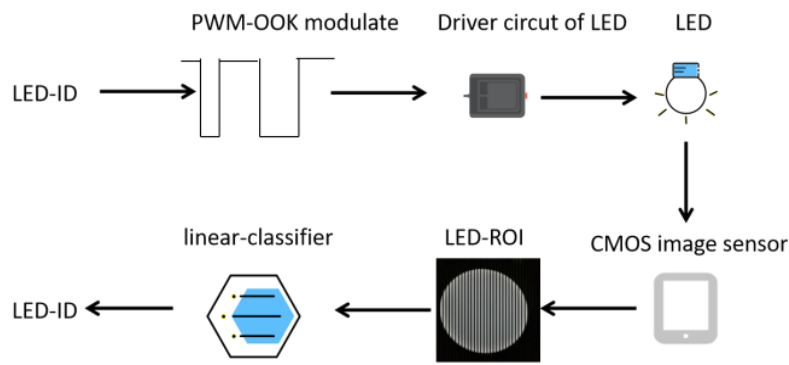

**Fig. 1.** The processing of Led-ID modulation and recognition caption.

Unlike the traditional LED-ID optical stripe modulation and demodulation algorithm based on VLC imaging communication, the purpose of modulation coding in this experiment is not to decode, but to give certain features to the light and dark stripes captured in the image. By the PWM-OOK modulization technology and Manchester code, we introduce four characteristic variables including frequency, duty cycle, distance, and phase difference coefficient to give features to the LED-ID optical stripe-code captured by the CMOS image sensors. The features are normally the number of stripes of the light stripe code in the LED pixel area, the area of the LED pixel area, the ratio of the width of the bright stripe to the width of the bright stripe and the dark stripe and phase difference coefficient between stripes (color feature). These features are later extracted by a simple image processing technology to recognize the location information of the LED-ID through the preestablished database.

## 2.1.2. IMPROVED TARGET SIGNAL SOURCE TRACKING AND EXTRACTION METHOD

In the previous chapters, we have known that the success of LED-ID detection and recognition method are inseparable from the accurate detection of Light-emitting diode region of interest(LED-ROI), and the detection effect of that determines the real-time performance and robustness of the system. However, traditional LED-ROI detection method based on pixel strength requires a large amount of computation due to the thorough detection of each frame of picture, resulting in a decline in real-time performance. In addition, if the LED is occluded or influenced by other non-signal sources, it may fail to detect the LED-ROI. In order to solve these problems, we have proposed a VLC dynamic location tracking detection method based on improved camshift-kalman algorithm.

The Camshift algorithm is improved Meanshift algorithm, which realizes target detection based on the distribution state of color probability density. It is widely used in the application field of target tracking due to its characteristics of no parameters, high speed and efficiency. However, the tracking accuracy of it will be greatly reduced in the case of similar background color interference. As a result, Kalman filter is applied to adjust the final result. Kalman filtering is an optimal estimation, which is widely used in the motion-state estimation of the targets. It has good tracking performance for the objects which moves at a constant speed, but may fail to maintain a good tracking effect when the motion state of the moving target changes suddenly. Therefore, in this experiment, an improved Camshift-Kalman algorithm is proposed.

The principle of the improved camshift-kalman algorithm is shown in Fig. 2. After initialization of the search window, the center and the size of the window is updated continuously through Meanshift algorithm until the moving distance of search window is less than the given threshold [29]. The output of the algorithm is considered as the measurement signal. And Bhattacharyya coefficient, the similarity factor between the measurement results of Camshift algorithm and the real target, is used to replace the measurement noise of Kalman filter. Then, Kalman filter is applied to correct the target position, which uses the state equation of linear system to estimate the optimal state of the system. In each image frame, the Bhattacharyya Coefficient between color histogram obtained by Camshift algorithm and the that of the real target is normalized as measured measurement noise in Kalman filter. Therefore, the variance of the whole measurement noise is time-varying, achieving real-time updating of the model.

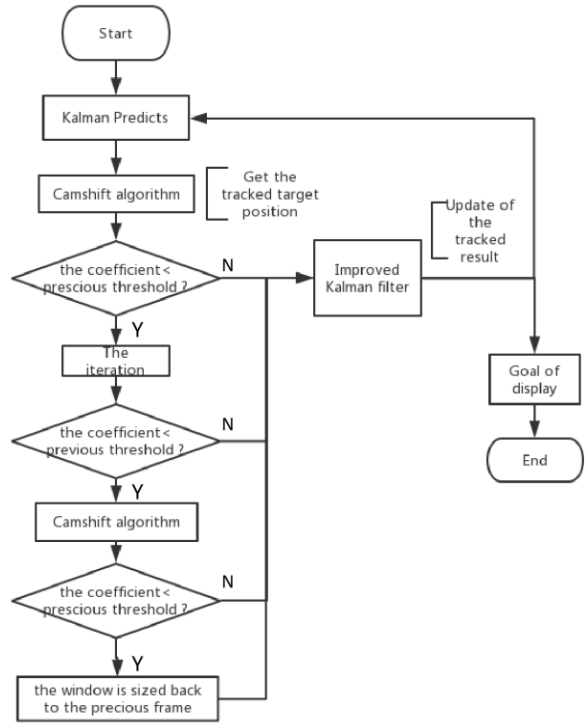

**Fig. 2.** The flow chart of Improved Target Signal Source Tracking and Extraction Method.

In some cases, when the coefficient is less than the preset threshold or when the target completely occludes for a period of time and then reenters the screen at a distance, some improvements are made such as moving the research window to eight adjacent areas to find the highest Bhattacharyya coefficient or continuously expanded the search window until the Bhattacharyya Coefficient is larger than the preset threshold. Then, the output research window is taken as the initial search window of next frame. By analyzing the video sequence obtained by image sensor, LED-ROI can be estimated after being detected in the initial image frame.

**2.1.3. THE DOUBLE-LIGHT POSITIONING ALGORITHM**

Generally, visible light positioning first builds an indoor environment map to match the LED-ID with the location coordinate on the map. Through the method mentioned in 2.1.1 and 2.1.2, the position of the LEDs in environment map can be got. After that, the location of the terminal relative to the position of the LEDs can be worked out. Due to the rolling shutter mechanism, when a CMOS camera images the LEDs, the image of LEDs are formed with light and dark stripes, the width of which depends on the flashing frequency of the LEDs. So, different lamps have distinct stripe features in the image [3]. If the camera moves, the position of the LED on the photo will also change. Therefore, the location of the camera can be determined based on the acquired LED-ID and the positions of each detected LEDs in the image. In the experiment, we adopted the camera imaging principle in computer vision to solve the problem of position with an industrial camera. As shown in Fig. 3, the LED is installed on the ceiling, whose coordinates are (x1, y1, z1), (x2, y2, z2). Point P is the midpoint of the lens in the image sensor and (i , j) is the coordinate of LEDs through the Lens in the image coordinate system. The focal length of the lens f is the intrinsic parameter of the camera.

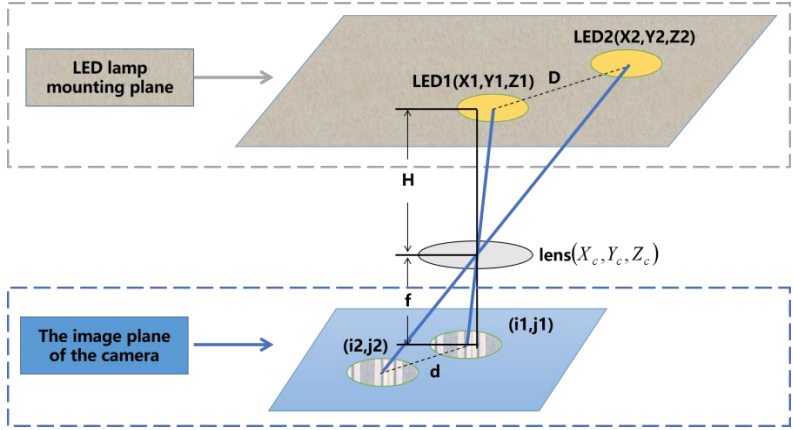

**Fig. 3.** Double-light positioning system model.

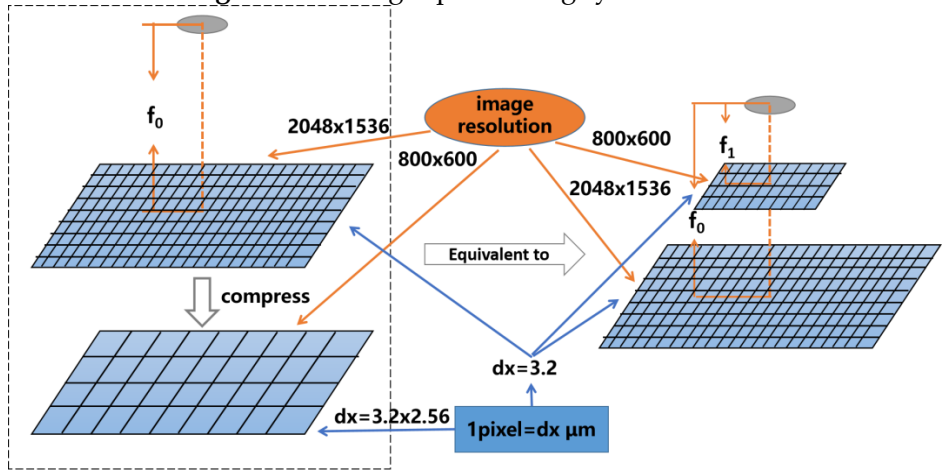

**Fig. 4.** Schematic diagram of equivalent change of camera focal length

In the process of visible light positioning, the problem of coordinate transformation should be considered. Although the pixel coordinate system and image coordinate system are both located on the imaging plane of image sensor, they have different origin and different units of measurement. So, the coordinate in pixel coordinate system is changed into the coordinate in image coordinate system [21]. The vertical distance from the image sensor to the led can be obtained according to the coordinates of the led.

$$H = f\frac{D}{d} = f\frac{\sqrt{(x_1-x_2)^2+(y_1-y_2)^2}}{\sqrt{(i_1-i_2)^2+(j_1-j_2)^2}} \quad (1)$$

As shown in Fig. 11, in experiments E and F, uncompressed images were used with a pixel size of 2048×1536. In this case, the focal length f0 of the camera was directly derived from the supplier of the camera. However, in experiments A,B,C and D, the image was compressed to 800×600. In principle, the focal length of the camera is constant, while the physical size corresponding to a pixel should be increased by a(a=2.56=2048/800) times. The physical size of the image is constant. In order to simplify the calculation, we use the concept of "equivalent focal length", which is different from the usual equivalent focal length. We reduce f0 by a times, so that the corresponding physical size of a pixel remains the same. In other words, the compressed focal length of the image is f1=f0/a. Formula (2) represents formula 1 before image compression, while formula (3) represents formula (1) after image compression.

$$\frac{D}{d_0} = \frac{H}{f_0}, d_0 = \Delta d_0 \times dx \tag{2}$$

$$\frac{D}{d_1} = \frac{H}{f_1}, d_1 = \Delta d_1 \times dx, \Delta d_0 = \Delta d_1 \times a \tag{3}$$

Where d represents the physical distance between the two lights in the image plane, $\Delta d$ represents the pixel distance between the two lights, a is the scaling factor of the picture and dx represents the conversion from pixel distance to physical distance.

If the image coordinate system and the world coordinate system are parallel and they have the same direction, the coordinate of the terminal (x, y) in the camera system can be calculated according to the principle of similar triangles:

$$\frac{x - \frac{x_1 + x_2}{2}}{H} = \frac{\frac{i_1 + i_2}{2}}{f} \tag{4}$$

$$\frac{y - \frac{y_1 + y_2}{2}}{H} = \frac{\frac{j_1 + j_2}{2}}{f} \tag{5}$$

However, in the actual situation, the image coordinate system is inconsistent with the coordinate axis direction of the world coordinate system, that is, there exist a rotation angle $\theta$ of the Zc axis. Therefore, the rotation angle of the image sensor is calculated with respect to the Zc axis of the world coordinate system. It is assumed that the vector of the centroid of the source LED1 to the centroid of the source LED2 is parallel to the axis Xc, and the directions are the same. So, the rotation angle $\theta$ can be calculated by the following formula:

$$\theta = \operatorname{atan2}\left(j_1 - \frac{j_1 - j_2}{2}, i_1 - \frac{i_1 - i_2}{2}\right) = \operatorname{atan2}(j_1 - j_2, i_1 - i_2) \tag{6}$$

However, in reality, the coordinate directions of the image coordinate system and the world coordinate system are not always consistent, so we solve the problem by calculating the rotation angle of the image sensor relative to the Zc axis in the world coordinate system. In our previous work in [32], the true rotation angle named $\varphi$ is calculated. Then, we can rotate the coordinate by the formula (7) and realize positioning.

$$\begin{bmatrix} x_w \\ y_w \\ z_w \end{bmatrix} = \begin{bmatrix} \cos\varphi & \sin\varphi & 0 \\ -\sin\varphi & \cos\varphi & 0 \\ 0 & 0 & 1 \end{bmatrix} \begin{bmatrix} x \\ y \\ z \end{bmatrix} \tag{7}$$

(X, Y, Z) is the coordinate in the image sensor coordinate system and (Xw, Yw, Zw) is the coordinate in the world coordinate system. Through formula (7), we can get the value of the coordinate of the terminal $(x_w, y_w)$.

**2.2. DESIGN OF VLC POSITIONING PACKAGE BASED ON ROS SYSTEM**

ROS system is an open source system for robots. It is a framework applicable to robot programming, which provides a communication framework for robots. Although ROS is called an operating system, it is not an operating system in the common sense like Windows or Mac. It just connects the operating system with the developed ROS application. Therefore, it is also a runtime environment on Linux, where the sensing, decision-making and control algorithms of the robot can be better organized and run. One of its advantages is that it can facilitate reuse of code in the development of robot program. The working process of it can be generated as packages or stacks, making it more convenient to finish the repetitive work.

The VLC package is a ROS package developed by ourself, whose structure is shown in Fig. 5. The package is divided into four nodes: 'mvcam', 'CamKF', 'ID_recognition' and 'vlc_locator'. All nodes can be run in two computers. The details are described below as Fig. 5:

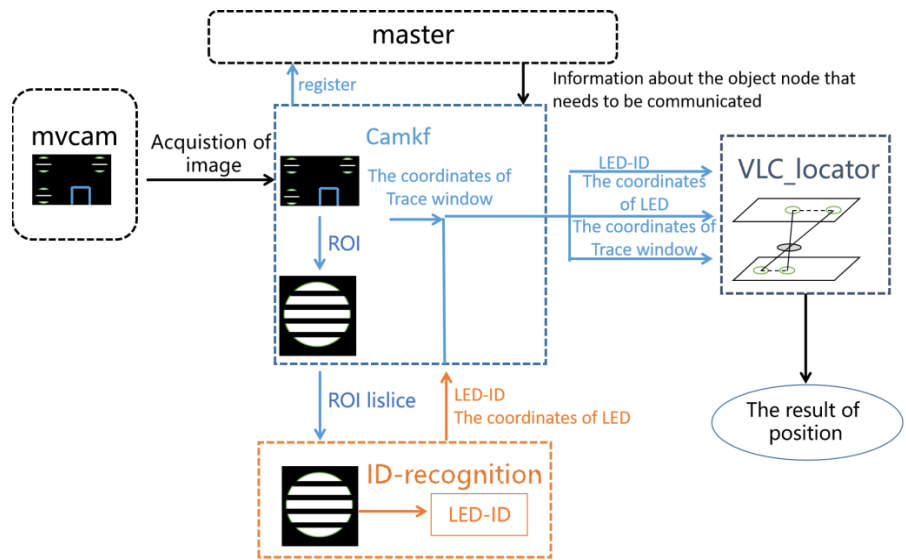

**Fig. 5.** VLC package structure.

### 2.2.1. The design of mvcam node

The node of "mvcam" means "MindVision camera node" —the CMOS industrial camera installed on the turtlebot3. It receives the signal from the LED, converts the images to ROS image messages and then publishes the message of LED image to the topic. The pseudo-code of the mvcam node is as follows:

| Node 1 | mvcan |
|---|---|
| **1:** | **Step 1.** Set the camera |
| **2:** | CameraSdkInit(1);//Camera initialization |
| **3:** | CameraSetAeState (hCamera,FALSE); //Set to custom mode |
| **4:** | CameraSetAnalogGain (hCamera,100); //Set the analog gain to 100 |
| **5:** | CameraSetExposureTime (hCamera,200);//Set the exposure time to 200 microseconds |
| **6:** | CameraSetIspOutFormat(hCamera,CAMERA_MEDIA_TYPE_MONO8);//Set the image output format to 8-bit gray |
| **7:** | CameraSetMonochrome (hCamera,TRUE); //Enable the color to black and white function |
| **8:** | CameraSetFrameSpeed (hCamera,2000); //Set camera output image frame rate to very high. |
| **9:** | Set the Camera parameters; |
| **10:** | **Step 2.** Get the image and publish it |
| **11:** | image_transport::Publisher pub =it.advertise("camera/image", 1); //Set the image publisher and the topic is camera/image |
| **12:** | ros::Rate r(fps); |
| **13:** | while (nh.ok()) |
| **14:** | { |
| **15:** |   CameraGetImageBuffer(hCamera,&sFrameInfo,&pbyBuffer,1000) ; //Gets a frame of image data. |
| **16:** |   CameraImageProcess(hCamera, pbyBuffer, g_pRgbBuffer,&sFrameInfo); //Process images according to preset parameters. |
| **17:** | resize(cvarrToMat(iplImage),frame,Size(800,600),0,0,INTER_NEAREST); |
| | //Scale the image to 800×600 pixels |
| **18:** | sensor_msgs::ImagePtr msg = cv_bridge::CvImage(std_msgs::Header(), "bgr8", frame).toImageMsg(); |
| | //Convert images from OpenCV format to ROS image messages via CvBridge |
| **19:** | pub.publish(msg); //publish a message |
| **20:** | ros::spinOnce(); |
| **21:** | r.sleep(); |
| **22:** | } |

### 2.2.2. The design of CamKF node

This node receives messages from the mvcam node, getting LED-ROI based on the Improved Target Signal Source Tracking and Extraction Method which is mentioned above. A structure is defined in this node, whose members include Vector tracking window, luminaire ID and x and y coordinates of the tracking window in the picture. During the tracking process, the ID information of the lamps in the current image and the coordinates of the tracking window (that is, the coordinates of the lamps in the image) are encapsulated in the customized message, and the positioning calculation request is issued to the positioning computing node through the service, and the message is sent to the positioning computing node. If a new unmarked luminaire appears in the image in the subsequent process, the ID identification request is immediately issued to the ID identification node through the service. The pseudo-code of CamKF node is as follows:

| | |
|---|---|
| **Node 2** | **CamKF** |
| **1:** | Struct LED A, B, C ; //Three lamps A,B and C are defined by the structure; |
| **2:** | Int main //Image processing main function |
| **3:** | { |
| | image_sub_ = it_.subscribe("/camera/image", 1, & IMAGE_LISTENER_and_LOCATOR::convert_callback, this); //Define the image receiver |
| **4:** | ros::Rate loop_rate(); |
| **5:** | while (ros::ok()) |
| **6:** | { |
| **7:** | { |
| **8:** | cv_ptr = cv_bridge::toCvCopy(msg, sensor_msgs::image_encodings::RGB8); |
| **9:** | While |
| | { |
| **10:** | ROI_process ; // traversing all ROI, looking for ROI; |
| **11:** | ros::ServiceClient client = nh.serviceClient< ID_recognition:: ID_recognition >("get_image"); |
| **12:** | ID_recognition:: ID_recognition ID_recognition_srv; |
| | // Declare a service called ID_recognition_srv that USES the ID_recognition service file |
| **13:** | ID_recognition_srv.request::sensor_msgs::ImagePtr msg = cv_bridge::CvImage(std_msgs::Header(), "bgr8", frame).toImageMsg(); //OpenCV image conversion to ROS image format |
| **14:** | srv.response.ID = ID; // get the ID by responding |
| **15:** | } |
| **16:** | } |
| **17:** | Tracking |
| | { |
| **18:** | ros::ServiceClient ros_tutorials_service_client =nh.serviceClient<LED_info:: LED_info >("LED_info_srv"); |
| **19:** | LED_info:: LED_info LED_info _srv; // declare a service called LED_info_srv that USES the LED_info service file |
| **20:** | Tracking(frame, trackObjectNum, frameIndex); // track lamp image, calculate tracking window |
| **21:** | LED_info_srv.request.A.ID = A.ID |
| **22:** | } |
| **23:** | ros::spin(); |
| **24:** | } |
| **25:** | } |

### 2.2.3. The design of ID_recognition node

This node responds to the ID identification request of the node 'CamKF', using the LED-ID detection and recognition method to recognize the image in LED-ROI and return the results to the node' CamKF'. The pseudo-code of the node is as follows:

| | Node 3 | ID_recognition |
|---|---|---|
| 1: | | bool get_image(ID_recognition::rgbd_image::Request &req, ID_recognition::ID::Response &res) |
| 2: | | // service response function |
| 3: | | { |
| | |   ID = Image_process（req. rgbd_image); |
| 4: | |   res.ID=ID ; // take the ID information as the service response |
| 5: | |   ROS_INFO("success"); |
| 6: | |   return true; |
| 7: | | } |
| 8: | | int main |
| 9: | | { |
| 10: | |   ros::ServiceServer service = n.advertiseService("get_image", get_image);// define service server "get_image" |
| | | //Process the image and get the ID |
| 11: | |   ros::Rate loop_rate(); |
| 12: | |   while (ros::ok()) |
| 13: | |   { |
| 14: | |    ros::spin();// waiting for service request |
| 15: | |   } |
| 16: | | } |

### 2.2.4. The design of vlc_locator node

The package of "vlc_locator" is the image processing and positioning computing node. It subscribes ROS image messages from topic and then processes the image to get the position. They communicate with each other through a topic in ROS. The communications between nodes are performed by the XML/RPC calls under the TCP/IP protocol. When the VLC program starts, we run the camera node on the TurtleBot and then run the locator node on remote controller. The camera node publish image in the topic, and the locator node subscribe to the topic to receive image, and then use the image to computing the position of TurtleBot. The pseudo-code of the vlc_locator node is as follows:

| | Node 4 | vlc _locator |
|---|---|---|
| 1: | | Struct LED A, B, C ; |
| | | //Three lamps A,B and C are defined by the structure; |
| 2: | | bool position(ros_tutorials_service::SrvTutorial::Request &req, ros_tutorials_service::SrvTutorial::Response &res) |
| 3: | | { |
| 4: | |   Req.A.ID=A.ID |
| 5: | |   Req. A_ Img_local_X =A. A_ Img_local_X |
| 6: | |   ROS_INFO("sending back response", (int)res.result);// response information |
| 7: | |   return true; |
| | | } |
| 8: | | ros::ServiceServer service = nh.advertiseService("LED_info _srv", position); |
| | | // define the service server and execute the position function. |
| 9: | | Get_coordinate(LED A, LED B, LED C) // coordinates are calculated according to the position of the luminaire. |
| 10: | | { |
| 11: | |   Global_coordinate_judgment (LED A, LED B, LED C);//Choose two lamps with different X and Y coordinates. |
| | |   //Assign to LED U1, LED U2. |
| 12: | |   pose_value= Calculate (LED U1, LED U2); |
| 13: | |   return pose_value; |
| 14: | | } |
| 15: | | int main() // locate the main function |
| 16: | | { |
| 17: | |   ros::Publisher pub = nh_.advertise ("location", 1000); // define a location publisher |
| 18: | |   { |
| 19: | |    ros::Rate loop_rate(); |
| 20: | |    while (ros::ok()) |

```
21:                   {
22:                       pose_value=Get_coordinate(LED A, LED B, LED C);// positioning according to lamp information
23:                       pub.publish(msg); // publish location information;
24:                   }
25:              ros::spin();
26:              }
27:          }
```

## 3. Experiment
### 3.1. Experiment setup

The setup of experiment is shown at tableⅠ. The images of LEDs were shot by MindVision UB-300 industrial camera and transmitted by Raspberry Pi 3 Model B with Quad ARM Cortex-A53 Core 1.2GHz Broadcom BCM2837 with 64bit CPU and 1GB RAM [35]. Because the Raspberry Pi 3 Model B is the Single Board Computers of TurtleBot3 Burger, which has weak processor performance, it is not suitable for image processing and positioning calculations to be expanded on TurtleBot. Based on distributed design of the ROS platform, image processing and positioning computing programs can be easily run on remote controller. The receiving and processing experiments were performed on remote controller Acer VN7-593G with i7-7700HQ, 3.40 GHz clock rate (Actual Turbo Frequency) and 16G RAM. The OS of the TurtleBot 3 is Ubuntu MATE 16.04, and the OS of remote controller is Ubuntu 16.04 desktop. In the experiment, the TurtleBot and remote controller are connected to the same WiFi, making it easy to get online in the Local Area Network (LAN). To verify the feasibility of the proposed indoor navigation system, our experiments were performed on an experimental platform with 4 LEDs which shown as Fig. 6 (Since the robot cannot observe lamps on the far side at the edge, we used three or four LED luminaires in the experiment to locate the entire area so that the robot can observe at least two LED luminaires at various locations) . It is set up to be 1-meter long, 1-meter wide and 1.5-meter high, and the available area for VLC positioning is a square with a side length of 1 meter. The experiment used distortion-free lenses, which allowed us to obtain better imaging results without additional camera correction.

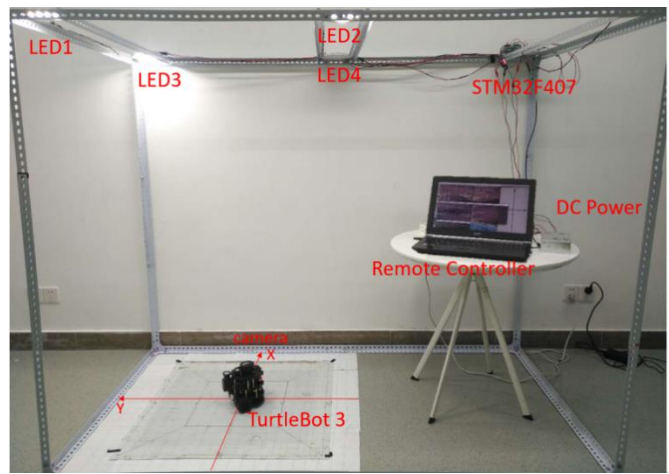

**Fig. 6.** Experimental platform of ROS based VLP System.

As described in the section System Principle, when the VLC program starts, we run the camera node on the TurtleBot and then run the locator node on remote controller. Afterwards, the camera node publishes image in the topic, and the locator node subscribe to the topic to receive image, and then use the image to compute the position of TurtleBot.

**Table 1.** Parameters Of Experimental Facilities And Platform.

| Camera Specifications | |
|---|---|
| Model | MindVision UB-300 |
| Spectral Response Range/nm | 400~1030 |
| Pixel(H × V) | 2048 × 1536 |
| Resolution | 800 × 600 (in experience A, B C and D) |
| | 2048 × 1536 (in experience E and F) |
| Pixel Size/μm2 | 3.2 × 3.2 |
| Time of Exposure/ms | 0.02 |
| Type of Shutter Acquisition Mode | Electronic Rolling Shutter |
| Acquisition Mode | Successive and Soft Trigger |
| Working Temperatures/°C | 0–50 |
| **Turtlebot3 Robot Specifications** | |
| Module | Raspberry Pi 3 B |
| CPU | Quad Core 1.2 GHz Broadcom BCM2837 |
| RAM | 1 GB |
| operating system | Ubuntu mate 16.04 |
| **Remote Controller Specifications** | |
| Module | Acer VN7-593G |
| CPU | Quad Core Intel® Core™ i7-7700HQ |
| RAM | 16 GB |
| operating system | Ubuntu 16.04 |
| **Experimental Platform Specifications** | |
| Size (L ×W× H)/ cm³ | 100 × 100 × 150 |
| **LED Specifications** | |
| Coordinates (x, y, z)/cm | LED1(-465,495,150) |
| | LED2(-460,-420,150) |
| | LED3(460,490,150) |
| | LED4(480,-425,150) |
| The half-power angles of LED/deg($\psi$1/2) | 60 |
| **Circuit Board Specifications** | |
| Drive chip | DD311 |
| Drive current/A | 0.1 |
| Drive voltage/V | 27 |

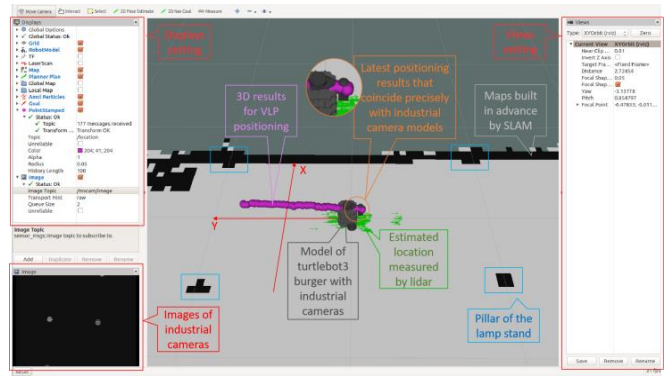

**Fig. 7.** User Interface of the proposed ROS robot VLP localization.

In the experiment, we used a 3d visualization platform-rviz, which displays a variety of data, as shown in Fig. 7. The display setting is used to control the displayed information, and the perspective setting is used to set the viewing angle. The image shown at the bottom left is that of an industrial camera. The middle part of the image is the main interface of the robot visualization, showing the current map and the VLP position estimation results. The map was constructed by SLAM and indicated the locations of the four columns and walls corresponding to the experimental scene map. The purple dots in the figure represent the positioning results obtained by using the VLP, while the green arrow indirectly reflects the position of the robot obtained by the lidar alone. The latest position estimation results can be seen in the local close-up, which is exactly in line with the position of the industrial camera model on the robot. Among them, the robot model with industrial camera represents the position estimation obtained by AMCL fusion localization by lidar and IMU.

### 3.2. Experiment and Analysis of Results
### 3.2.1. Stationary Positioning

The stationary coordinate experiment considers the stationary robot to be located at coordinate (0,0), and then obtain results of the positioned coordinate. As the robot is stationary, the state of the robot remains unchanged during the time process. Because the size of the platform is large, the robot cannot be located precisely. So, the dispersion radius is used to characterize the positioning accuracy while calculating the accuracy of positioning. Since the vast majority of positioning results are distributed in a small area, the size of the dispersion area can characterize the density of data distribution. The smaller the dispersion circle, the denser the data distribution, the higher the accuracy. The diameter of the dispersion circle can be defined as the farthest distance of the two points in all the results, then vast majority of the experimental results will be concentrated in this circle. Ideally, if the center of the dispersion circle coincides with the coordinates of the origin after correction, the positioning error can be represented as the Euclidean distance between the data and the center of the dispersion circle. If the dispersion radius is small enough, accurate results can be obtained after precise calibration of the camera.

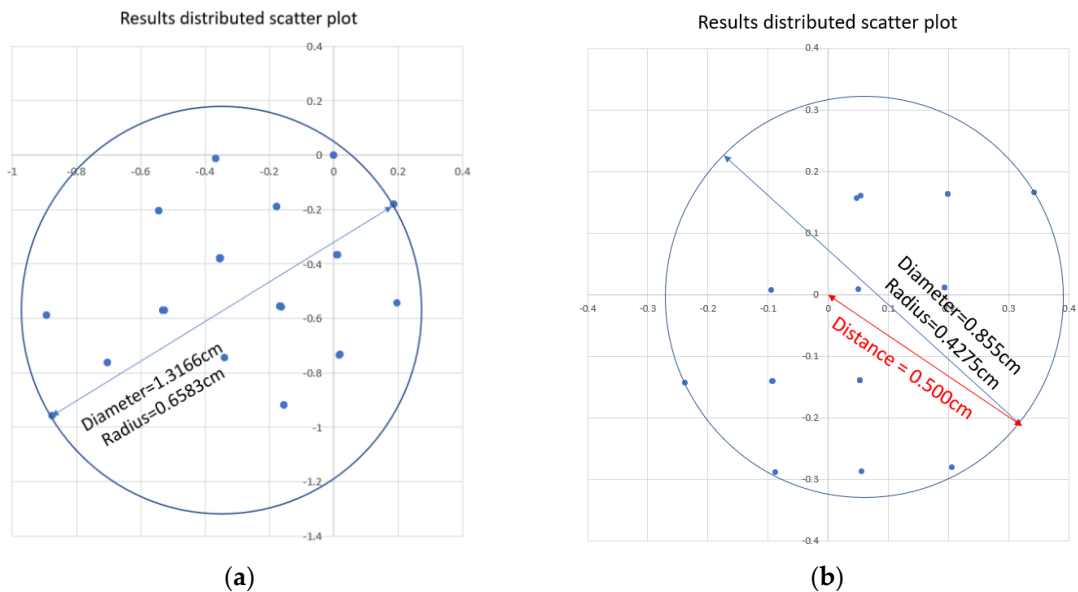

(a) (b)

**Fig. 8.** Estimated location of uncorrected fixed position positioning experiments.

As is shown in Fig. 8(a), the actual position of the robot may deviate from the origin, but the dispersion radius is only 0.6583 cm. This means that the positioning error would be 0.6583 cm in the ideal state.

In this experiment, we were lucky enough to find a way to substantially reduce this error: we measured the position of each lamp as precisely as possible, fixed the camera to the robot in the most robust way, and calculated the actual image center point coordinates by static positioning experiment results. The following is a comparison of the results, such as Fig. 8(b), the dispersion radius is only 0.4275 cm, the center of the circle is only 0.8cm from the origin, and the furthest distance from the origin is only 0.5cm.

To verify the accuracy of static positioning, we have 36 uniformly distributed measuring points at 0 height. Each measuring point was measured 12 times, resulting in a total of 432 data. The test steps for each point are as follows: the coordinate paper provides precise coordinates, and then the moving robot moves to the corresponding position of the coordinate paper, comparing the positioning results with the actual location coordinates of the point, resulting in an error.

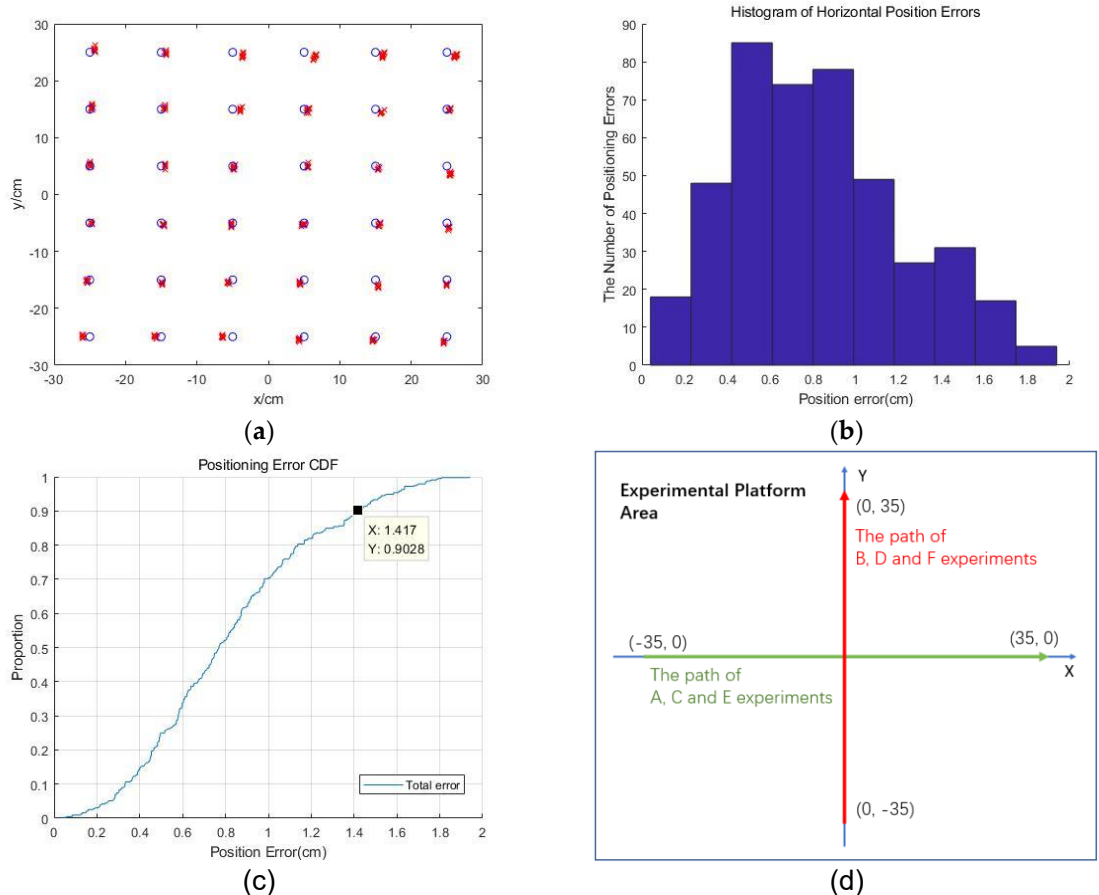

**Fig. 9.** Static positioning results and precision.(a) Estimated position of the corrected fixed position positioning experiment.(b) The probability mass function (PMF) curves of positioning error in the real world.(c) The cumulative distribution function (CDF) curves of positioning error in the real world.(d)The path of four mobile positioning experiments.

As shown in Fig. 9, the 90% positioning error is less than 1.417cm, the average positioning error is 0.82cm. This approach will be discussed in more detail in another paper to be published, and this article provides a brief introduction to this approach and its effects.

**3.2.2. Dynamic Positioning**

As shown in Fig. 9(d),in the mobile positioning experiments, the robot moves along the given path, then the VLP system gets the estimated position coordinates. In the experiment, we accurately measure the starting and ending coordinates of the robot's movement, draw the robot's actual moving trajectory in a straight line, and calculate the error on this basis. In addition, we will give the robot moving trajectory line obtained by the fit of the experimental data, and show both lines in the result graph for visual evaluation of the overall error level and trend of positioning.

### 3.2.2.a. Positioning accuracy analysis

As shown in Fig. 9(d), in A and B experiments, the robot connected with a camera moves along the longitudinal and transverse centerline of the experimental area at a speed of 0.4 cm/s respectively. The experimental results are shown in the scatter graph, fitting line diagram, probability mass function (PMF) and cumulative distribution function (CDF) of the positioning results.

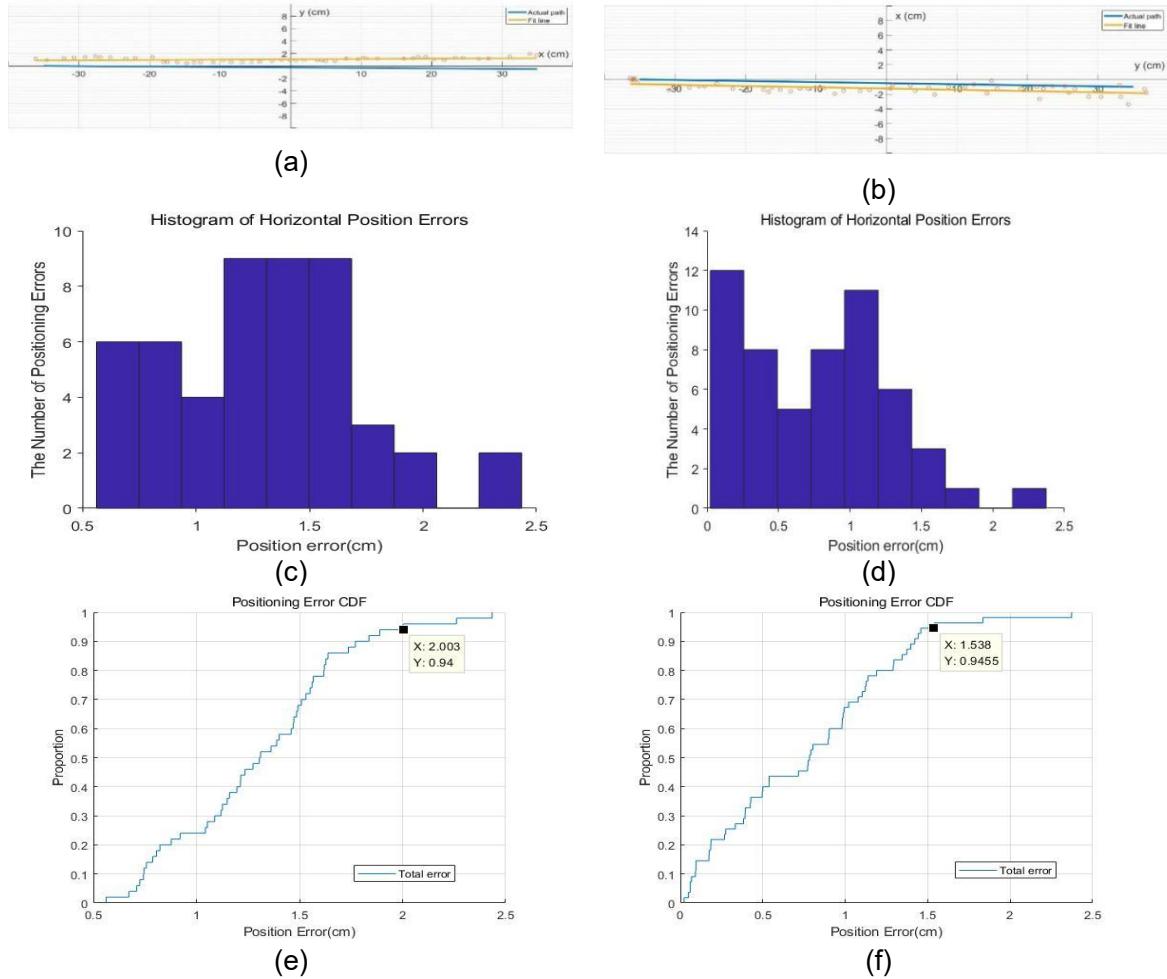

(a)

(b)

(c)

(d)

(e)

(f)

**Fig. 10.** Estimated position and positioning error of the A experiment, TurtleBot3 move from (-35, 0) to (35,-0.5), the camera connect to the TurtleBot3, resolution of the images is 800×600 pixels.(a) Estimated position distributed scatter plot and fitting straight line.(c) The probability mass function (PMF) curves of positioning error in the real world.(e) The cumulative distribution function (CDF) curves of positioning error in the real world. Estimated position and positioning error of the B experiment, TurtleBot3 move from (0, -35) to (-1, 35), the camera connect to the TurtleBot3, resolution of the images is 800×600 pixels.(b) Estimated position distributed scatter plot and fitting straight line.(d) The probability mass function (PMF) curves of positioning error in the real world.(f) The cumulative distribution function (CDF) curves of positioning error in the real world.

The fitting linear equation of A experiment is:

$$y=0.0051x+1.0714 \tag{8}$$

And the fitting linear equation of B experiment is:

$$x=-0.0171y-1.2286 \tag{9}$$

### 3.2.2.b. Real-Time Performance

Due to the poor performance of the Raspberry Pi, it takes a few seconds to calculate a result, so that we must use distributed computing in this experiment. Although distributed computing is one of the advantages of ROS, it can lead to serious problems in this experiment. In the process of publishing images on Raspberry Pi and subscribing it on remote controller, the ROS publish-subscribe framework has a big impact on the transfer performance of the subscribed images across devices. The original image, which is taken by the industrial camera in the experiment platform, is 2048×1536 pixels. When an image of this size is subscribed through ROS, it takes two seconds to transfer the image to another device. After testing, we find images below 800×600 pixels can get better real-time performance in cross-device transmission, making the distributed positioning program reach the acceptable level. However, this effect does not exist if the image is published and subscribed in the same device.

As shown in Fig. 11, much less stripes can be recognized from the image of 800x600 pixels compared with the image of 2048x1536 pixels, which means less luminaire IDs can be accommodated in the same hardware environment. In order to explore the performance of a camera when it is connected directly to a computer, C and D experiments were designed, in which 800×600 pixels images were used in C experiment and uncompressed 2048x1536 pixels images were used in D experiment. And the two nodes were tested in Acer VN7-593G with i7-7700HQ, which is a high-performance laptop. The experimental results show smaller stripes can be identified in the images, and position results are calculated faster than distributed computing.

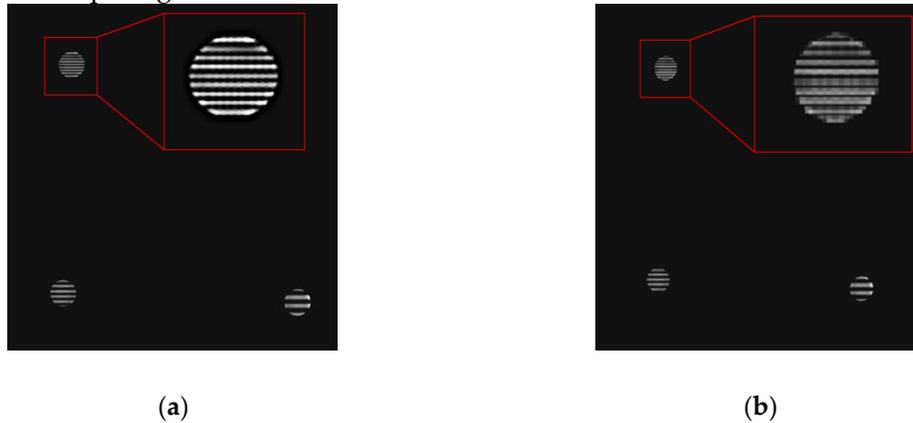

(a)          (b)

**Fig. 11.** Image comparison at different resolutions, (a) uncompressed 2048×1536 pixels images used in E, F experiment, (b) compressed 800×600 pixels images used in A, B, C and D experiment.

In C, D, E and F experiments, the camera is connected to the remote controller, and the robot is moving along the longitudinal centerline of the experimental area in the experiments. In all the four experiments, the robot's speed was 0.4 cm/s.

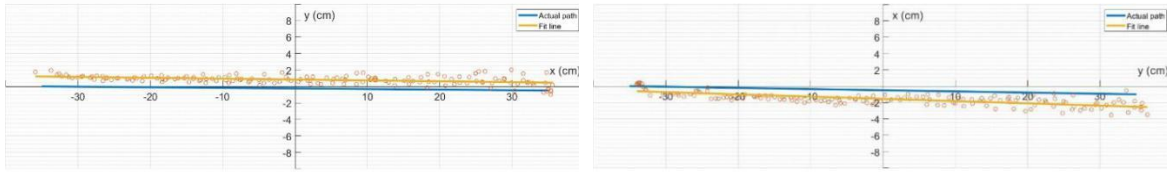

(a)                      (b)

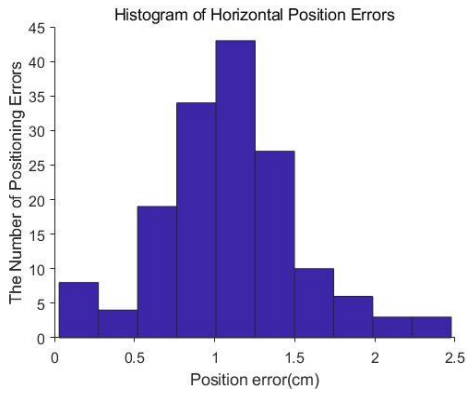 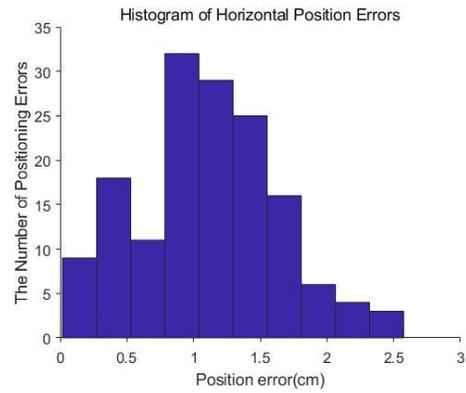

(c)                      (d)

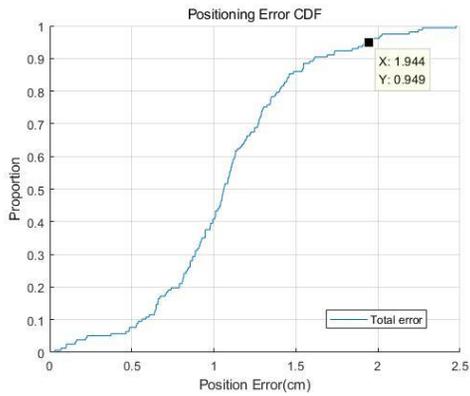 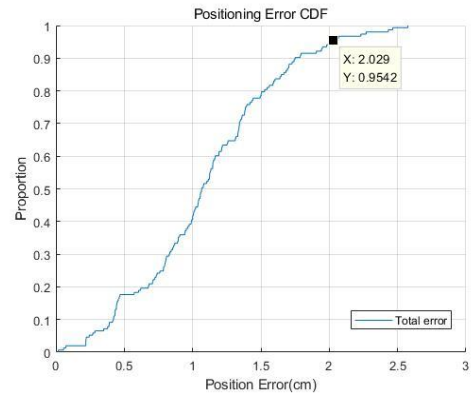

(e)                      (f)

**Fig. 12.** Estimated position and positioning error of the C experiment, TurtleBot3 move from (-35, 0) to (35, -0.5), the camera connect to the remote controller, resolution of the images is 800×600 pixels. (a) Estimated position distributed scatter plot and fitting straight line.(c) The probability mass function (PMF) curves of positioning error in the real world. (e) The cumulative distribution function (CDF) curves of positioning error in the real world. Estimated position and positioning error of the D experiment, TurtleBot3 move from (0, -35) to (-1, 35), the camera connect to the remote controller, resolution of the images is 800×600 pixels.(b) Estimated position distributed scatter plot and fitting straight line.(d) The probability mass function (PMF) curves of positioning error in the real world.(f) The cumulative distribution function (CDF) curves of positioning error in the real world.

The fitting linear equation of C experiment is:

$$y=-0.01058x+0.8293 \tag{10}$$

And the fitting linear equation of D experiment is:

$$x=-0.0275y-1.5530 \tag{11}$$

From Fig. 12 (c), (e) and Fig. 12 (d), (f) we can learn that more than 95% of positioning error were less than 1.944 cm and 2.029 cm while the robot moves along the transverse and the longitudinal centerline of the experimental area respectively.

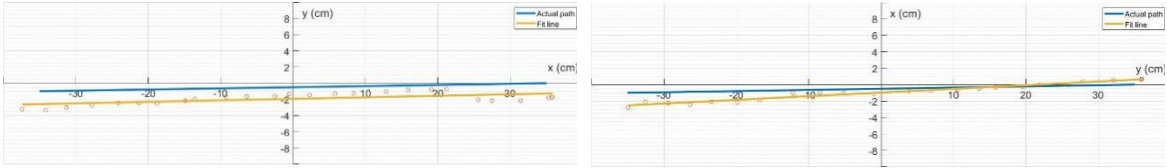

(**a**)                      (**b**)

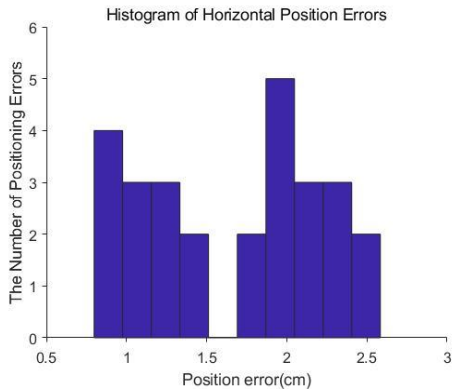
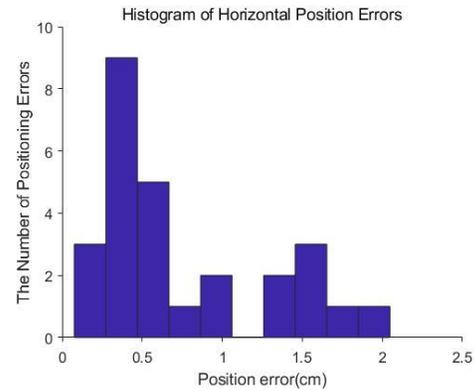

(**c**)                      (**d**)

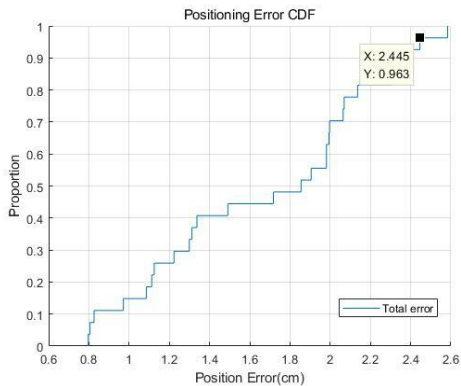
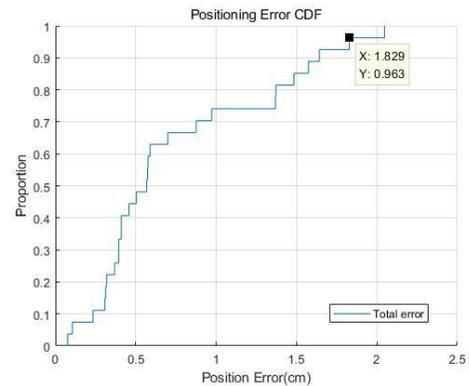

(**e**)                      (**f**)

Fig. 13 Estimated position and positioning error of the E experiment, TurtleBot move from (35, 0) to (-35, -1), resolution of the images is 2048×1536 pixels. the camera connect to the remote controller.(a) Estimated position distributed scatter plot and fitting straight line.(c) The probability mass function (PMF) curves of positioning error in the real world.(e) The cumulative distribution function (CDF) curves of positioning error in the real

world. Estimated position and positioning error of the F experiment, TurtleBot move from (0, 35) to (-1, -35), resolution of the images is 2048×1536 pixels. the camera connect to the remote controller.(b) Estimated position distributed scatter plot and fitting straight line.(d) The probability mass function (PMF) curves of positioning error in the real world.(f) The cumulative distribution function (CDF) curves of positioning error in the real world.

The fitting linear equation of E experiment is:

$$y=0.0187x-1.9313 \tag{12}$$

And the fitting linear equation of F experiment is:

$$x=0.0445y-0.9643 \tag{13}$$

From Fig. 13 (c), (e) and Fig. 13 (d), (f) we can learn that more than 96% of positioning error were less than 2.445 cm and 1.829 cm while the robot moves along the transverse and the longitudinal centerline of the experimental area respectively.

**Table 2.** Average positioning time of four experiment.

| Experiment | (A) | (B) | (C) | (D) | (E) | (F) |
|---|---|---|---|---|---|---|
| Average time (s) | 0.3619 | 0.3538 | 0.1207 | 0.1221 | 0.7970 | 0.8085 |

Real-time performance is reflected in the timestamp of each result. The average positioning time based on timestamps is shown in the table Ⅱ. As it shown, the average positioning time of experiment C and D is significantly shorter, and the average positioning time of experiment E and F was two times longer than that of A and B when 2048x1536 pixel images were used in the experiment. The processing capacity of Turtlebot3 is not enough to handle the VLC positioning algorithm proposed in this paper, so it needs to be handled by remote controller. In summary, it is recommended to use a high-performance computer on the robot, at least the Intel® Core™ U-Series Processor, to run the camera node and the locator node together. Otherwise, other measures need to be taken to avoid the delay caused by the transmission of images across devices and the loss of resolution resulting from compromise on real-time requirements.

However, with the improvement of the processing capacity of Turtlebot, the VLC positioning algorithm proposed in this paper will be processed directly on the robot. So, only the positioning results will be fed back to the remote control book, which can greatly improve the real-time performance of the system.

## 4. Conclusions

In order to apply visible lighting position to ROS, a real-time indoor three-dimensional visible light positioning (VLP) system based on ROS and TurtleBot3 robot platform is proposed. Based on the loose coupling characteristics of the ROS system, the positioning algorithm is implemented by loosely coupled ROS nodes. The images from industrial camera are published by the camera node and subscribed by the locator node, then the position results are computed. Experimental results show that the proposed system can provide indoor localization within 2 cm and possesses a good real-time performance which takes only 0.35 seconds for one-time positioning. And if a high-performance laptop is used, the single positioning time can be reduced to 0.12 seconds.

More tests and improvements are needed in the future to improve the comprehensive performance of the software system. For example, it can be deeply integrated into the positioning system of the TurtleBot and the corresponding user interface based on the Rviz software development. Besides, it is also beneficial to the study of ROS Fusion localization. Because of the inherent shortcomings of visible light communication, using it alone will inevitably lead to insufficient robustness. On the other hand, as the experiment shows, the robot can not walk accurately as the given straight line. So, it needs feedback from other external positioning methods to form a closed-loop control.

Therefore, in the future work we can build a system combined with visible light communication and multiple sensors which includes inertial navigation system to increase the robustness, the positioning accuracy and the real-time ability of the positioning system.

**Acknowledgments**

This work was supported in part by the National Undergraduate Innovative and Entrepreneurial Training Program under Grants 201510561003, 201610561065, 201610561068, 201710561006, 201710561054, 201710561057, 201710561058, 201710561199, and 201710561202; in part by the Special Funds for the Cultivation of Guangdong College Students' Scientific and Technological Innovation ("Climbing Program" Special Funds) under Grants pdjh2017b0040 and pdjha0028; and in part by the Guangdong Science and Technology Project under Grant 2017B010114001.